\documentstyle[12pt,epsf,epsfig]{article}
\newcommand{\al}{{\alpha}}

\newcommand{\btab}{\begin{tabbing}}
\newcommand{\etab}{\end{tabbing}}

\newcommand{\beqn}{\begin{equation}}
\newcommand{\eeqn}{\end{equation}}
\newcommand{\barr}[1]{\begin{array}{#1}}
\newcommand{\earr}{\end{array}}
\newcommand{\beqna}{\begin{eqnarray}}
\newcommand{\eeqna}{\end{eqnarray}}
\newcommand{\btablec}{\begin{table} \begin{center}}
\newcommand{\etablec}{\end{center} \end{table}}

\newcommand{\gapproxeq}{\lower.7ex\hbox{$\;\stackrel{\textstyle>}
{\sim}\;$}}
\newcommand{\plabel}[1]{\label{#1}}
\newcommand{\pbibitem}[1]{\bibitem{#1}}
\def\question#1{}
\input epsf

\begin{document}
\title{
\begin{flushright} 
\small{nucl-th/0506063} \\ 
\small{LA-UR-05-4415} 
\end{flushright} 
\vspace{0.5cm}  
\Large\bf New broad $^8$Be nuclear resonances}
\vskip 0.2 in
\author{Philip R. Page\thanks{\small \em E-mail:
prp@lanl.gov}\\{\small \em  Theoretical Division, MS B283, Los Alamos
National Laboratory, Los Alamos, NM 87545, USA}}
\date{}
\maketitle
\begin{abstract}
Energies, total and partial widths, 
and reduced width amplitudes of $^8$Be resonances up to an
excitation energy of $26$ MeV are extracted from a coupled channel
analysis of experimental data. The presence of an extremely broad
$J^\pi=2^+$ ``intruder'' resonance 
is confirmed, while a new $1^+$ and very broad $4^+$ resonance
are discovered. A previously known $22$ MeV $2^+$ resonance is 
likely resolved
into two resonances. The experimental $J^\pi T=3^{(+)}?$ resonance at
$22$ MeV is determined to be $3^{-}0$, and the experimental
$1^-?$ (at $19$ MeV) and $4^-?$ resonances to be isospin 0.
\end{abstract}
\bigskip

Keywords: $^8$Be, resonance, R-matrix

PACS number(s): 27.20.+n \hspace{.16cm} 25.10.+s
\hspace{.16cm}  21.10.Dr \hspace{.16cm} 21.10.Hw
\hspace{.16cm}  21.10.Jx

\vspace{.3cm}

\section{Introduction \label{sec1}}

What are the properties of the resonances of $^8$Be? This question is
most comprehensively answered by a global analysis of all experimental
data based on the best reaction theory available, for example R-matrix
theory. Resonance structure tends to be based on single experiments,
most recently compiled by TUNL~\cite{tunl}. In contrast, the results of a
coupled channel R-matrix analysis of data from 69 experimental
references are given here. This analysis does not include all
experimental data, and hence is not expected to provide the best
parameters for all resonances. This is particularly true of narrow
resonances (with widths less than 100 keV), which can be approximated
by the Breit-Wigner formula. The strength of a coupled channel
R-matrix analysis becomes apparent for broad resonances, whose
structure can only be determined by analyzing data over a large energy
range in various channels, and for which the full force of reaction
theory is needed.

The physical content of scattering can be summarized by knowledge of
the S-matrix for real energies. However, a more intuitive picture is
provided by resonances, which are defined as complex energy poles of
the S-matrix. The real part of the pole $\lambda$ is defined as the
excitation energy $E_x$, and two times the imaginary part as the
width $\Gamma$. Because these parameters can only be found for
complex energies, which cannot be experimentally accessed, resonances
involve a mathematical extrapolation beyond observation. Since
resonances with small widths tend to have the most pronounced
experimental effects, this analysis is limited to resonances fairly
near to the real energy axis (the 
``unphysical sheet closest to the physical sheet''~\cite{hale87}). Even so,
controversy centers around very broad resonances which are not
observable as clear bumps in experimental cross-sections, particularly
a total angular momentum, parity, isospin and excitation energy
$J^\pi T(E_x)=2^+0(16)$
resonance found in this analysis. This
resonance was previously found in an R-matrix analysis by Barker {\it
et al.}~\cite{barkerold,barker88,barker89,barker00}.  
They also found a broad $0^+$ at about
$10$ MeV. 
This analysis also discovers a previously unreported broad
$4^+0(18)$ resonance.

\section{Analysis technique~\label{sec2}}

The analysis is performed with the EDA R-matrix
code~\cite{eda}. Integrated cross-section, differential cross-section
and polarization data, consisting of more than
4700 points, are fitted with a
$\chi^2/(d.o.f.)$ of $7.4$ utilizing about $100$ free parameters
(the R-matrix level eigenenergies and reduced width amplitudes discussed in the
next section). This high $\chi^2$
is mostly related to contradictory data, as well as underestimates of
experimental relative and normalization errors~\cite{laur}. Since the
resonance structure is insensitive to exclusion of data that fit with
more than three standard deviations~\cite{laur}, it is robust under
inclusion of the worst fitting data points. Experimental nuclear data
on the reactions $^4He(\alpha,\alpha_0)$, $^4He(\alpha,p_0)$,
$^4He(\alpha,d_0)$, $^7Li (p,\alpha_0)$, $^7Li (p,p_0)$, $^7Li (p,n_0)$, $^7Be
(n,p_0)$, $^6Li (d,\alpha_0)$, $^6Li (d,p_0)$, $^6Li (d,n_0)$ and $^6Li
(d,d_0)$, leading to the $^8$Be intermediate state, are included. 
All recoil nuclei are in the ground state. 
Table~\ref{tablong} contains a complete list of the data in the analysis.
Substantial data are entered
for the $^4He(\alpha,\alpha_0)$ and $^7Li (p,p_0)$ reactions, and the
least data are entered for the $^4He(\alpha,p_0)$, $^4He(\alpha,d_0)$ and
$^6Li (d,d_0)$ reactions~\cite{laur}. The maximum excitation energy
above the $^8$Be ground state is $25-26$ MeV for all reactions except
$^4He(\alpha,\alpha_0)$ and $^7Be (n,p_0)$. In the $^4He(\alpha,\alpha_0)$
reaction, data above the maximum $\alpha$ laboratory energy for which
data are entered ($38.4$ MeV) and below the limit of this analysis,
are only available as phase shifts~\cite{bacher}, and have not been
incorporated.  For the $^7Be (n,p_0)$ reaction no data above the
near-threshold data entered are found below the maximum excitation
energy of this analysis. Further details of the data and cross-section
fits are available~\cite{laur,nd2004}.

The excitation energies of the thresholds of the various analyzed
channels, with respect to the unstable $^8$Be ground state, are
$-0.09$ ($\alpha\: ^4$He), $17.26$ ($p\: ^7$Li), $18.90$ ($n\: ^7$Be)
and $22.28$ MeV ($d\: ^6$Li)~\cite{tunl}. The two-body channels $p\:
^7$Li$^\ast$, $n\: ^7$Be$^\ast$ and $d\: ^6$Li$^\ast$, involving
resonances less than $100$ keV wide, are neglected.  These could
reasonably be included in an R-matrix analysis.  All the channels
included are strongly constrained by unitarity (via the R-matrix
formalism) and, as explained in the next section,
isospin symmetry (charge independence). 
The channel radii are fixed as follows based on earlier
R-matrix analyses: $\alpha\: ^4$He (4.0 fm), $p\: ^7$Li and $n\:
^7$Be (3.0 fm) and $d\: ^6$Li (6.5 fm). The fit is insensitive to
variation in the $d\: ^6$Li radius~\cite{laur}.  The orbital angular
momenta included between the two scattered nuclei are: $\alpha\: ^4$He
(S-~, D-~, G-~, I- and L-waves), $p\: ^7$Li and $n\: ^7$Be (S-~, P-~,
D-~ and F-waves) and $d\: ^6$Li (S-~, P-~ and D-waves). The inclusion
of the highest wave for each channel did not seem to change the
qualitative features of the fit, indicating that a sufficient number
of waves has been used.

\begin{table}
\begin{center}
\begin{tabular}{|l||l|c||l|c|}
\hline 
Reaction & Data Reference & $E$ (MeV) & Data Reference & $E$ (MeV) \\ 
\hline \hline 
$^4He(\alpha,\alpha)$ & Heydenburg 1956~\cite{heydenburg} & $0.6-3.0$ & Phillips 1955~\cite{phillips} & $3.0-5.8$ \\
& Tombrello 1963~\cite{tombrello} & $3.8-11.9$ & Steigert 1953~\cite{steigert} & $12.9-20.4$ \\
& Chien 1974~\cite{chien} & $18.0-29.5$ & Mather 1951~\cite{mather} & $20.0$ \\
& Nilson 1956~\cite{nilson} & $12.3-22.9$ & Briggs 1953~\cite{briggs} & $21.8-22.9$ \\
& Bredin 1959~\cite{bredin} & $23.1-38.4$ & Graves 1951~\cite{graves} & $30.0$ \\
\hline 
$^4He(\alpha,p)$ & King 1977~\cite{king} & $39.0-49.5$ & & \\
\hline 
$^4He(\alpha,d)$ & King 1977~\cite{king} & $46.7-49.5$ & & \\
\hline 
$^7Li(p,\alpha)$ & Spraker 2000~\cite{spraker} & $0.0-0.1$ & Harmon 1989~\cite{harmon} & $0.0-0.3$ \\
& Rolfs 1986~\cite{rolfs} & $0.0-1.0$ & Engstler 1992~\cite{engstler} & $0.0-1.3$ \\
& Cassagnou 1962~\cite{cassagnou} & $1.4-4.8$ & Kilian 1969~\cite{kilian} & $3.4-9.4$ \\
& Freeman 1958~\cite{freeman} & $1.0-1.5$ & Mani 1964~\cite{mani} & $3.0-10.1$ \\
\hline 
$^7Li (p,p)$ & Warters 1953~\cite{warters} & $0.4-1.4$ & Bardolle 1966~\cite{bardolle} & $0.8-2.0$ \\
& Lerner 1969~\cite{lerner} & $1.4$ & Malmberg 1956~\cite{malmberg} & $1.3-3.0$ \\
& Gleyvod 1965~\cite{gleyvod} & $2.5-4.2$ & Brown 1973~\cite{brown} & $0.7-2.4$ \\
& Bingham 1971~\cite{bingham} & $6.9$ & Kilian 1969~\cite{kilian} & $3.1-10.3$ \\
\hline 
$^7Li (p,n)$ & Macklin 1958~\cite{gibbons,macklin} & $1.9-3.0$ & Barr 1978~\cite{barr} & $2.0-3.0$ \\
& Burke 1974~\cite{burke} & $1.9-3.0$ & Meadows 1972~\cite{meadows} & $1.9-3.0$ \\
& Elbakr 1972~\cite{elbakr} & $2.2-5.5$ & Darden 1961~\cite{darden} & $2.0-2.3$ \\
& Austin 1961~\cite{austin} & $2.1-3.0$ &  Elwyn 1961~\cite{elwyn62} & $2.0-2.6$ \\
& Baicker 1960~\cite{baicker} & $3.0$ & Andress 1965~\cite{andress} & $3.0$ \\
& Hardekopf 1971~\cite{hardekopf} & $3.0$ & Thornton 1971~\cite{thornton} & $3.0-5.5$ \\ 
& Poppe 1976~\cite{poppe} & $4.3-10.0$ & & \\
\hline 
$^7Be (n,p)$ & Koehler 1988~\cite{koehler} & $0.0-0.0$ & Cervena 1989~\cite{cervena} & $0.0$ \\
\hline 
$^6Li (d,\alpha)$ & Engstler 1992~\cite{engstler} & $0.0-1.3$ & Golovkov 1981~\cite{golovkov} & $0.1-0.1$ \\
& Elwyn 1977~\cite{elwyn} & $0.1-1.0$ & Bertrand 1968~\cite{bertrand} & $0.3-1.0$ \\
& Cai 1985~\cite{cai} & $0.5-2.5$ & McClenahan 1975~\cite{mcclenahan} & $0.5-3.4$ \\
& Jeronymo 1962~\cite{jeronymo} & $0.9-5.0$ & Gould 1975~\cite{gould} & $2.2-4.9$ \\
& Risler 1977~\cite{risler} & $1.0-5.0$ & & \\
\hline 
\end{tabular}
\end{center}
\end{table}

\begin{table}
\begin{center}
\begin{tabular}{|l||l|c||l|c|}
\hline 
Reaction & Data Reference & $E$ (MeV) & Data Reference & $E$ (MeV) \\ 
\hline \hline 
$^6Li (d,p)$ & Szabo 1982~\cite{szabo82} & $0.1-0.2$ & Body 1979~\cite{body} & $0.1-0.2$ \\
& Bertrand 1968~\cite{bertrand} & $0.3-1.0$ & Elwyn 1977~\cite{elwyn} & $0.1-1.0$ \\
& Cai 1985~\cite{cai} & $0.5-2.5$ & McClenahan 1975~\cite{mcclenahan} & $0.5-3.4$ \\
& Bruno 1966~\cite{bruno} & $1.0-2.0$ & Gould 1975~\cite{gould} & $2.3-5.0$ \\
& Durr 1968~\cite{durr} & $2.1-4.8$ & & \\
\hline 
$^6Li (d,n)$ & Hirst 1954~\cite{hirst} & $0.1-0.3$ & McClenahan 1975~\cite{mcclenahan} & $0.5-2.9$ \\
& Szabo 1977~\cite{szabo77} & $0.1-0.2$ & Haouat 1985~\cite{haouat} & $0.2-1.0$ \\
& Elwyn 1977~\cite{elwyn} & $0.2-0.9$ & Bochkarev 1994~\cite{bochkarev} & $0.8$ \\
& Thomason 1970~\cite{thomason} & $2.5-3.7$ & & \\
\hline 
$^6Li (d,d)$ & Abramovich 1976~\cite{abramovich} & $3.0-5.0$ & & \\
\hline 
\end{tabular}
\caption{\plabel{tablong} Data in the $^8Be$ analysis. The laboratory energy of the
projectile is $E$.}
\end{center}
\end{table}

\section{Procedure}

The Kapur-Peierls expression for the S-matrix at real energies $E$ for
channels $c'$ and $c$ is (Eq. 28 of Ref.~\cite{lane})

\beqn\plabel{sm}
S_{c'c} = \frac{I_c(a_c,k_c)}{O_c(a_c,k_c)}\delta_{c'c} + i \sum_\mu
\frac{\rho_{\mu c'} \rho_{\mu c}}{{\cal E}_\mu(E) - E}
\hspace{1cm}\mbox{where  } \rho_{\mu c} = 
\frac{\sqrt{2 k_c a_c} {\cal G}_{\mu c}(E)}{O_c(a_c,k_c)}  .
\eeqn
Here the incoming and outgoing wave functions $I$ and $O$ are functions
of $E$ through the wave number $k$. In principle 
the S-matrix is independent of
the channel radii $a$. The complex functions ${\cal E}_\mu(E)$ and
${\cal G}_{\mu c}(E)$ are determined by the R-matrix
fit (see below, and also Ref.~\cite{lane}).
Eq.~\ref{sm} can be extended
to complex $E$, and the S-matrix remains independent of $a$.
The poles of the S-matrix then occurs
at complex $E_0={\cal E}_\mu(E_0)$, where 
$E_x\equiv Re[E_0]$ is the resonance excitation energy 
and 
$\Gamma \equiv -2\, Im[E_0]$ is 
the resonance total width. The partial width 
$\Gamma_c\equiv |\rho_{\mu c}|^2=
2\,|k_{0c}|\,a_c\,|{\cal G}_{\mu c}(E_0)/O_c(a_c,k_{0c})|^2$ 
is evaluated at the pole in terms of the reduced width amplitude
$g_c\equiv|{\cal G}_{\mu c}(E_0)|$, and
is related to the residue at the pole (see Eq.~\ref{sm}). 
The quantities $E_x$, $\Gamma$ and $\Gamma_c$ are independent of $a$.
Contrary to physical intuition, 
the sum of $\Gamma_c$ for kinematically open channels
is {\it not} equal to $\Gamma$.
It should be cautioned that $E_x, \; \Gamma$ and
$\Gamma_c$ all depend on how the extension to complex $E$
is done, and are accordingly quantities that cannot be
measured experimentally. However, for narrow resonances where
${\cal E}_\mu(E)$ is almost real, $E_x, \; \Gamma$ and
$\Gamma_c$ respectively 
collapse to the usual notions of excitation energy,
width and partial width, which can be measured experimentally. 

 The method of calculation of the S-matrix poles and residues in terms of
the R-matrix parameters is briefly summarized from the more complete
discussion~\cite{hale87}. To obtain the S-matrix pole positions
from the real R-matrix eigenenergies $E_\lambda$ and the real reduced width
amplitudes $\gamma_{\lambda c}$ for the real boundary conditions 
$B_c$ (fixed in this analysis), as defined in Ref.~\cite{thomas}, a
complex energy $E_0$ is found such that at least one eigenvalue of the complex
``energy-level'' matrix (p. 294 of Ref.~\cite{thomas}),

\beqn\plabel{elm}
{\cal E}_{\lambda' \lambda} \equiv E_\lambda \delta_{\lambda' \lambda}
-\sum_c \gamma_{\lambda' c} [L_c(a_c,k_c) -B_c] \gamma_{\lambda c}
\eeqn
is the same as $E_0$. Here the outgoing-wave logarithmic
derivatives $L$ are defined in terms of the outgoing wave functions
$O$ in the usual way (Eq. 4.4, p. 271 of Ref.~\cite{thomas}),
and are functions of $E$ through the wave number $k$.
The residue at the pole $i\rho_{\mu c'}\rho_{\mu c}$ 
has already been written in terms of the function ${\cal G}_{\mu c}(E_0)$ 
in Eq.~\ref{sm}. This function can be calculated from the
R-matrix parameters by using Eq. 4 of Ref.~\cite{hale87}.
Although this function and the energy-level matrix (Eq.~\ref{elm})
are defined for real energies, extension to complex $E$ is done by simply
using the functional form of these expressions when working with complex
energies. 
In this way both the S-matrix pole $E_0$ and the function 
${\cal G}_{\mu c}(E_0)$, needed to calculate the excitation energy, 
(partial) width
and reduced width amplitude, are defined in terms of the R-matrix parameters.

The EDA code~\cite{eda} used to perform the R-matrix analysis
implements the standard Wigner R-matrix theory~\cite{thomas} without
approximations, except for restricting the number of R-matrix levels
for a given $J^\pi T$ to a finite number of levels in the energy
region of interest.  The analysis
employs isospin symmetry in the limited sense that isospin constraints
on the $\gamma_{\lambda c}$ are implemented as follows.  The $\alpha\:
^4$He and $d\: ^6$Li channels couple to an isospin 0 level, but not to
an isospin 1 level. Hence the $\gamma$'s for an isospin 1
level coupling to these channels are set to zero.  Also, a level's
$\gamma$'s for the $p\: ^7$Li and $n\: ^7$Be channels are
related by isospin Clebsch-Gordon coefficients, which are different for
isospin 0 and 1 levels.

Let us consider the dissociation of the compound nucleus $A$ into
nucleus $A'$ and ejectile $a$. Define the channel cluster form factor
$F$, proportional to the overlap between the internal wave function of
nucleus $A$ and the internal wave functions of the nuclei $A'$ and
$a$, as~\cite{ichi}

\beqn
F(r_{aA'}) \sim \int [\psi_{A'}(\xi_{A'})
\psi_a(\xi_a)]^\ast \psi_A(\xi_A) d\xi_{A'}d\xi_a\; .
\eeqn
Here $r_{aA'}$ is the relative coordinate between the C.M. of $a$ and
$A'$. The symbols $\xi_A,\; \xi_{A'}$ and $\xi_a$ denote internal coordinates
of the nuclei $A$, $A'$ and $a$, respectively; and $\psi$ are 
the corresponding
internal wave functions. A full definition of $F$ can be found elsewhere 
(Eq. 7 of Ref.~\cite{navratil04}). The integral of $|F|^2$ over $r_{aA'}$ 
is the widely predicted ``spectroscopic factor''.
The R-matrix reduced width amplitude $\gamma_{\lambda c}$ for the 
breakup of a level $\lambda$ 
of the nucleus $A$ into $A'$ and $a$ in channel $c$ is defined
as~\cite{thomas,ichi}

\beqn\plabel{comp}
\gamma_{\lambda c} = \sqrt{\frac{\hbar^2 a_c}{2M_c}}\; F(a_c)\; ,
\eeqn
where $M_c$ is the reduced mass for relative motion between
$A'$ and $a$.
Comparison between theory calculations and the predictions here are
possible by comparing $F(a_c)$ calculated from theory
and $\gamma_{\lambda c}$ using
Eq.~\ref{comp}. However, this is only possible when the same boundary
conditions $B_c$ are imposed at $a_c$ is as standardly done in
R-matrix theory. As theory calculations do not usually do this, it is
more useful to compare them to ${\cal G}_{\mu c}(E)$ in Eq.~\ref{sm},
which is the equivalent of $\gamma_{\lambda c}$ for wave functions
with outgoing wave (Kapur-Peierls) boundary conditions
(Eq. 30 of Ref.~\cite{lane}). Hence the R.H.S. of Eq.~\ref{comp},
calculated from theory (usually) for bound states, should be compared
to the $g_{c}$ which will be tabulated in the next section
for scattering states. 

\section{Resonance structure~\label{sec3}}

\begin{table}
\begin{center}
\begin{tabular}{|l||r|c||r|c||r|l|}
\hline 
$J^\pi T$ & \multicolumn{2}{c||}{$E_x$ (MeV)} & 
\multicolumn{2}{c||}{$\Gamma$ (keV)} & $i$ & \\ \cline{2-5}
& R-m. & Exp. & R-m. & Exp.  & \% &  \\
\hline \hline 
$0^+0$ & [0.01]& 0         & [0.01]    & 0.00557(25)  &[0]& $\dag\ddag$ \\
       & 20.13 & 20.2      &   750     & 720(20)      & 0 & $\dag\ddag$ \\
$1^+ 0 $&18.17 & 18.150(4) &   140     & 138(6)       & 1  &$\dag\ddag$ \\ 
$1^+ 1 $&17.66 & 17.640(1) &    10     & 10.7(5)      & 0  &$\dag\ddag$ \\
$1^+ 1$& 20.45 & -         &   690     & -            & 0  &$\dag$      \\   
$2^+0$ &  2.77 & 3.03(1)   &  1200     & 1513(15)     &[0]& $\dag\ddag$ \\
       & 16.40 & -         & 19200     & -            & 0 & $\dag\amalg$\\
       & 20.10 & 20.1      &   680     & 880(20)      & 4 & $\dag\ddag$ \\
       & 22.09 & 22.2      &   590     &$\approx 800$ & 0 & $\dag\ddag$ \\
       & 22.78 & -         &  1670     & -            & 0 & $\P$\\
       & 23.25 & 25.2      &  2000     & -            & 0 & $\dag$\\
$3^+ 0 $&19.24 & 19.24     &   170     & 227(16)      &29 & $\dag\ddag$ \\
$3^+ 1 $&19.02 & 19.07     &   270     & 270(20)      &30 & $\dag\ddag$ \\
$4^+0$ & 11.57 & 11.35(15) &  4400     &$\approx 3500$& 0 & $\dag\ddag$ \\
       & 17.59 & -         &  7900     & -            & 0 & $\dag\amalg$\\
       & 24.35 & 25.5      &  4600     & broad        & 0 & $\dag$\\
$1^-0$ & 19.33 & 19.4      &   650     & $\approx$ 645&13 & \\
$2^-1$ & 18.92 & 18.91     &   120     & 122          & 2 & \\
$3^-0$ & 21.35 & 21.5      &   950     & 1000         & 0 & \\
$4^-0$ & 21.50 & 20.9      &  1060     & 1600(200)    & 0 & \\
\hline 
\end{tabular}
\caption{\plabel{tab1} 
Comparison of R-matrix 
and ``experimental''~\protect\cite{tunl} energies 
$E_x$ and widths $\Gamma$ of $^8$Be
resonances.  
Energies are relative to the experimentally determined $^8$Be ground state. 
The experimental error is indicated in brackets.
The isospin impurity $i$ of the squared amplitude
means that $1-i$ of the resonance is in the isospin $T$
indicated in column 1. Theory calculations: Confirmed ($\dag$) or
not confirmed ($\P$) by NCSM~\cite{navratil01}.
Confirmed ($\ddag$) or
not confirmed ($\amalg$) by GFMC~\cite{carlson}.
Quantities in square brackets are not accurately determined by this
analysis. For a discussion of the $1^-1(22)$ resonance see the text.
}
\end{center}
\end{table}

The $E_x$, $\Gamma$ and isospin impurity of the
resonances are displayed 
in Table~\ref{tab1}. All $J^\pi$  
are allowed, so that the $J^\pi$ is independently established
by the R-matrix analysis. Isospin 0 and 1 are allowed for all
resonances, because
these are the only isospins that can couple to the channels in 
this analysis if isospin symmetry is assumed. 
The resonances found in Table~\ref{tab1} should be compared to the
``experimental'' resonances believed to exist on the basis of a
summary of resonances found in experimental data and other
analyses~\protect\cite{tunl}. A comparison with experiment indicates
substantial agreement. Disagreements partially stem from the
difference between defining the energy and width from poles of
the S-matrix, as is done in the R-matrix analysis, 
and defining them from Breit-Wigner
formulae, as is often the case in experimental analyses. 
For example, agreement between the energy and width of the well-known
narrowest resonances ($J^\pi T(E_x) = 0^+0(0)$, $1^+0(18)$, $1^+1(18)$, 
$3^+0(19)$ and $3^+1(19)$)
is much better than those of the well-known
broadest resonances ($2^+0(3)$ and $4^+0(11)$).
However, the parameters of the $4^+0(11)$ resonance found
from $^4He(\alpha,\alpha)$ alone ($E_x=11.5(3)$ MeV,
$\Gamma=4000(400)$ keV)~\cite{tunl} are in perfect
agreement with this analysis. 
Since the R-matrix analysis contains more data than any known
analysis, the experimental masses and widths may well be in
doubt, although this is less likely for narrow
experimental resonances.

Except for the two very narrow experimental resonances 
$2^+(16.6;16.9)$
that are
not considered in the R-matrix fit because no data
are entered in their energy region, 
the following experimental
resonances are not found in the analysis: $4^+0(20)$, 
$(1,2)^-1(24)$, and three
resonances in the region $22-23$ MeV 
with unknown $J^\pi T$~\cite{tunl}. 
For the latter three resonances,
and $(1,2)^-1(24)$, the reason is that these resonances were 
observed in reactions other than those analyzed here~\cite{tunl}.
Of the reactions studied here, the $4^+0(20)$ resonance
is only non-negligibly observed in $^4He(\alpha,\alpha_0)$~\cite{tunl},
and data from the experimental reference~\cite{bacher} are
not included here. 

The narrow ground state $0^+0$ resonance parameters in Table~\ref{tab1} 
are not an improvement on experiment, since no low-energy
$^4 He(\alpha,\alpha_0)$ data are included at the same
excitation energy as the resonance energy. 
The experimental $J^\pi T = 1^-?$ at $19$ MeV~\cite{tunl}, 
and the $4^-?$~\cite{tunl},
are found to have isospin 0, having allowed for both isospins.

The quantum numbers of the peak at $21.5$ MeV in the $^7Li(p,n_0)$
reaction is experimentally thought to be $J=3$, with the parity
possibly positive~\cite{tunl,gibbons,li7pnold}.  Our fits prefers the
quantum numbers $J^\pi T = 3^-0$, having allowed for both parity and
both isospin possibilities. The new data
included~\cite{elbakr,thornton,poppe} hence updates the old
experimental parity assignment based on old
data~\cite{gibbons,li7pnold}. A positive parity assignment of the
$21.5$ MeV resonance is inconsistent with theory for the following
reason.  The only kinematically allowed decay channels analysed here
are to $p\: ^7$Li and $n\: ^7$Be. The NCSM predicts that the $3^+0$
and $3^+1$ resonances above the lowest-energy resonances with the same
quantum numbers have weak couplings to $p\: ^7$Li and $n\:
^7$Be~\cite{navratil04}.  The same is true for VMC if the $T=1$ $^8$Li
states are taken as a guide to the $T=1$ $^8$Be
states~\cite{wiringa}. The weak couplings to $p\: ^7$Li and $n\: ^7$Be
are not consistent with the need for the resonance here.

Two resonances with the same quantum numbers are found at $22-23$ MeV
in Table~\ref{tab1}. The $2^+0(23)$ 
resonance at $22.78$ MeV
fits the peak observed around 1 MeV $d$ laboratory energy in
the $^6Li(d,\alpha_0)$, $^6Li(d,p_0)$ and $^6Li(d,n_0)$ reactions. 
On the other hand, the $2^+0(22)$ resonance fits the peak at
around 6 MeV $p$ laboratory energy in the $^7Li(p,\alpha_0)$, and around
45 MeV $\alpha$ laboratory energy in the time-inverse $^4He(\alpha,p_0)$
reactions.  Although it is conceivable that all these peaks can be
fitted with just one $2^+0$ resonance, 
with the $d\: ^6Li$ threshold at $22.28$ MeV, 
the current fit clearly prefers two resonances. The lower mass resonance
is well established~\cite{tunl}. The existence of the
higher mass resonance only became apparent once $^6Li(d,X)$ data
above $\approx 1$ MeV $d$ laboratory energy were included, and hence 
does not contradict an analysis~\cite{czerski} of $^6Li(d,\alpha)$ data
below 1 MeV which only found the $2^+0(22)$.
The existence of two $2^+$ resonances at 
$21.5$ MeV and $22.5$ MeV were previously suggested 
by a qualitative analysis~\cite{presser} of the
$^7Li(p,n_1)$ and $^7Li(p,p_1)$ reactions not analyzed here,
in order to explain a broad dip in the $n_1$ yield at the same 
energy as a broad bump in the $p_1$ yield.
However, this analysis cannot be regarded as strong evidence for 
two $2^+0$ resonances. It is unclear whether two 
$2^+0$  resonances at $22-23$ MeV is confirmed
by NCSM theory calculations~\cite{navratil01}. This calculation
{\it does} find an extra $2^+0$ state at $14-21$ MeV, 
which is known as an ``intruder'' state because it 
does not appear in the na\"{\i}ve shell model. Whether this intruder
should be identified with the $2^+0(23)$ or with the extremely
broad $2^+0(16)$, discussed below, is unclear.

The $23.25$ MeV resonance found in the R-matrix analysis 
(Table~\ref{tab1}) is denoted by $2^+0(25)$. The reason is that when the
peak in $^6Li(d,\alpha_0)$ at a $d$ laboratory energy of $\approx 3.5$ MeV is
artificially enhanced by substantially decreasing the size of the
error bars, the resonance appears at $25.06$ MeV, in agreement with
experiment, with an unchanged width.
 
Most of the resonances found in the R-matrix analysis correspond to
resonances known experimentally. The exceptions are the extremely
broad $2^+0(16)$ and very broad $4^+?(18)$ resonances
(as well as the $1^+(20)$ discussed in the next paragraph).  The $2^+0(16)$
has previously been reported in an R-matrix analysis of $\alpha\:
^4$He elastic scattering, $^9Be(p,d)$ and $\beta$-delayed $2\alpha$
spectra from $^8$Li and $^8$B~\cite{barkerold,barker89,barker00} at
$\approx 9$ MeV~\cite{barker88,barker89,barker00}.  The energy, but
not the existence, of this level is dependent on the channel radius
used in the R-matrix fit~\cite{barker00,warburton}.  For example, an
analysis of $\beta$-delayed $2\alpha$ spectra from $^8$Li and $^8$B
together with $\ell=2$ $\alpha\: ^4$He phase shifts finds that $2^+$
intruder states below excitation energy $26$ MeV need not be
introduced~\cite{warburton}.
Although the S-matrix (and its poles and residues) are formally
independent of the chosen channel radii for infinitely many R-matrix
levels, actual analyses employ a finite number of levels, which can
lead to different energies for different channel radii.  In addition,
the energy of $2^+0$ varies by several MeV as new data are included,
consistent with the expectation that the energy should not be
particularly well constrained for a very broad resonance.  A NCSM
theory calculation finds the $2^+0$ and $4^+0$ intruders at $14-21$
and $20-26$ MeV respectively~\cite{navratil01}.  However, a recent
GFMC calculation finds no need to introduce extra $2^+$ or $4^+$
states below respectively $22$ and $19$ MeV~\cite{carlson}.  The
disagreement between NCSM and GFMC may be due to the large widths of
the intruder states (Table~\ref{tab1}), which imply substantial
variation in the energies extracted from these calculations which
treat all the states as bound.  Whether very broad states should be
seen in calculations that treat states as bound is debatable.
 
The current fit has a new $1^+1(20)$ resonance. Although it is not
listed in the standard experimental compilation~\cite{tunl}, it is
interesting to note that theory calculations predict such states: NCSM
predicts one $1^+0$ resonance and two $1^+1$ resonances at $20-22$
MeV~\cite{navratil01}, and GFMC one $1^+0$ at $\approx 19$
MeV~\cite{carlson}.  It is intriguing to note two coincidences between
this analysis and theory. (i) The NCSM predicts large couplings of a
$\approx 20.37$ MeV $1^+1$ state to $p\: ^7$Li and $n\: ^7$Be and not
to $d\: ^6$Li~\cite{navratil04}. The robust $1^+1(20)$ resonance seen
in this analysis is at $E_x=20.45$ MeV from Table~\ref{tab1}, with
strong couplings to $p\: ^7$Li and $n\: ^7$Be and not to $d\: ^6$Li
according to Table~\ref{tab2}.  (ii) Of the three $1^+$ resonances
predicted at $20-22$ MeV in NCSM, only the $\approx 20.37$ MeV $1^+1$
has large couplings to $p\: ^7$Li and $n\: ^7$Be, which are the only
kinematically open channels for decay, amongst the channels analysed
here~\cite{navratil04}. The same is true for VMC if the $T=1$ $^8$Li
states are taken as a guide to the $T=1$ $^8$Be
states~\cite{wiringa}.  This coincides with the finding here that
only one new $1^+$ state is needed, and that this state has isospin 1.



\begin{table}
\begin{center}
\begin{tabular}{|l|l|l|}
\hline 
$J^\pi T(E_x)$ & $\Gamma_c$ (keV) & $g_c$ $\times$ 100 ($\sqrt{\mbox{MeV}}$) \\ 
\hline \hline 
$0^+$ & \multicolumn{2}{l|}{$\al\: 1s\;\;\:  p\:3p\;\;\:  n\:3p\;\;\:  d\:5d\;\;\:  d\:1s$} \\
$0^+0(0)$ & [0.010] & [82] \\
$0^+0(20)$ & 550 40 120 & 24 25 53 86 61 \\ \hline
$1^+$ &  \multicolumn{2}{l|}{$p\:5p\;\;\:  p\:5f\;\;\:  p\:3p\;\;\:  
n\:5p\;\;\:  n\:5f\;\;\:  n\:3p\;\;\:  d\:5d\;\;\:  d\:3s\;\;\:  d\:3d$} \\
$1^+0(18)$ & 81 0.00008 60 & 91 4 78 \\ 
$1^+1(18)$ & 5 0.00008 6 & 65 27 69 \\
$1^+1(20)$ & 220 23 160 170 4 80 & 51 182 44 55 181 38 5 1 4 \\ \hline
$2^+$ &  \multicolumn{2}{l|}{$\al\:1d \;\;\: p\:5p \;\;\: p\:5f \;\;\: 
p\:3p  \;\;\: p\:3f \;\;\: n\:5p \;\;\: n\:5f \;\;\: n\:3p \;\;\: 
n\:3f \;\;\: d\:5s \;\;\: d\:5d \;\;\: d\:3d \;\;\: d\:1d$} \\
$2^+0(3)$ & 910 & 100 \\
$2^+0(16)$ & 1930 & 170 17 13 54 13 17 14 54 13 10 65 31 11 \\
$2^+0(20)$ & 170 130 20 130 0.06 140 2 100 0.02 & 14 43 213 43 11 59 180 50 21 21 25 45 25 \\
$2^+0(22)$ & 110 240 9 10 7 280 5 10 4 & 11 42 55 9 46 49 61 10 52 36 68 49 14\\
$2^+0(23)$ & 40 290 2 20 0.8 260 2 20 0.3 230 & 7 45 17 11 13 46 29 13 10 69 35 24 8\\
 & 70 30 4 & \\
$2^+0(25)$ & 70 930 20 20 4 880 20 30 2 50 40 & 9 79 61 10 25 80 79 14 24 29 22 18 5 \\ 
 & 30 2 & \\ \hline
$3^+$ &  \multicolumn{2}{l|}{$p\:5p\;\;\:  p\:5f\;\;\:  p\:3f\;\;\:  
n\:5p\;\;\:  n\:5f\;\;\:  n\:3f\;\;\:  d\:5d\;\;\:  d\:3d$} \\
$3^+0(19)$ & 130 0.07 0.4 7 0.001 0.009 & 56 24 57 98 43 131 \\
$3^+1(19)$ & 320 0.3 2 3 0.0001 0.0004 & 96 60 157 30 43 82 \\ \hline
$4^+$ &  \multicolumn{2}{l|}{$\al\:1g\;\;\:  p\:5f\;\;\:  
p\:3f\;\;\:  n\:5f\;\;\:  n\:3f\;\;\:  d\:5d$} \\
$4^+0(11)$ & 4000 & 135 17 28 17 27 0.6 \\
$4^+0(18)$ & 5300 2 4 & 135 33 50 33 50 3 \\ 
$4^+0(24)$ & 50 40 70 30 60 800 & 9 60 77 59 77 107 \\ \hline
$1^-$ &  \multicolumn{2}{l|}{$p\:5d\;\;\:  p\:3s\;\;\:  p\:3d\;\;\:  n\:5d\;\;\:  n\:3s\;\;\: 
n\:3d\;\;\:  d\:5p\;\;\:  d\:3p\;\;\:  d\:1p$} \\
$1^-(19)$ & 44 230 110 4 280 9 & 101 45 158 101 65 156 \\ \hline
$2^-$ &  \multicolumn{2}{l|}{$p\:5s\;\;\:  p\:5d\;\;\:  p\:3d\;\;\:  n\:5s\;\;\:  n\:5d\;\;\: 
n\:3d\;\;\:  d\:5p\;\;\:  d\:3p$} \\
$2^-(19)$ & 3 0.4 73 80 0.03 0.08 & 5 14 178 58 127 208 \\ \hline
$3^-$ &  \multicolumn{2}{l|}{$p\:5d\;\;\:  p\:3d\;\;\:  n\:5d\;\;\:  n\:3d\;\;\:  d\:5p$} \\
$3^-0(21)$ & 220 340 120 190 & 96 119 95 120 4 \\ \hline
$4^-$ &  \multicolumn{2}{l|}{$p\:5d\;\;\:  n\:5d$} \\
$4^-0(21)$ & 610 350 & 153 153 \\ 
\hline 
\end{tabular}
\end{center}
\end{table}

\begin{table}
\begin{center}
\caption{\plabel{tab2}
The partial widths $\Gamma_c$ and reduced width amplitudes 
$g_c$ found in the R-matrix analysis.  
First, the list of possible channels is indicated 
for each $J^\pi$. Each channel is denoted in the format (reaction)
$(2s+1)$ $\ell$;  where ``reaction'' is $\alpha$ ($\alpha\: ^4$He), 
$p$~($p\: ^7$Li), $n$~($n\: ^7$Be) or $d$~($d\: ^6$Li); and $s$ and $\ell$ are the
spin and orbital angular momentum of the nuclei in the 
channel. Second, for each resonance, $\Gamma_c$ and $g_c$
are indicated in the order of the channels enumerated for the corresponding
$J^\pi$. These entries always start with the first channel, but do not
necessarily end with the last channel. For
$\Gamma_c$ this is because the corresponding channels are
not kinematically allowed. For $g_c$ the quantities could 
not be determined because the resonance is too distant from the 
relevant threshold. Quantities in square 
brackets are not accurately determined by this analysis.
It is understood that $\Gamma_c$ and $g_c$ are only
given for the channels considered in this analysis; and that
certain two-body channels, all three-body channels, and higher $\ell$,
are neglected. The $g_c$ are channel radius dependent, and hence
not experimentally measurable.
}
\end{center}
\end{table}

The $2^-$ resonance is conceptually complicated because it lies
exactly at the $n\: ^7$Be threshold, and hence requires sophysticated
analysis. Several such analyses have been performed~\cite{tunl}, 
typically yielding a resonance with $E_x = 18.9$ MeV and $\Gamma 
\approx 100$ keV, although there is disagreement on the width.
Most strikingly, an analysis of $^7Li(p,n_0)$ and
$^7Be(n,p_0)$ data finds $\Gamma=1634$ keV~\cite{adah}, 
based on a prescription whereby the sum of the $\Gamma_c$
equals $\Gamma$. As previously mentioned, this is not the case in
our analysis. In contrast, another multi-level R-matrix 
analysis~\cite{koehler}
defines the resonance energy and width as the properties of the
pole of the S-matrix, yielding a total width much lower than the
sum of the partial widths. This
corresponds closely to our conventions, yielding $\Gamma=122$ 
keV, $T=0$ and isospin impurity $\approx 24$\%~\cite{koehler}.
This isospin impurity is at odds with $\leq 10$\% obtained from
$^7Li(p,\gamma)^8Be^\ast(18.9)$~\cite{tunl}. 
The current analysis assigns 
$T=1$ for the $2^-$ resonance (Table~\ref{tab1}). 
A cautionary note should be mentioned.
For all the resonances reported here except the $2^-$,
the parameters of the pole on the unphysical sheet closest to the 
physical sheet~\cite{hale87} are quoted in Tables~\ref{tab1}-\ref{tab2}, 
as this is thought to be physically most
relevant. However, there are poles on other sheets which are physically
less relevant. The $2^-0(19)$ is unique in that the resonance is
very close to threshold, which blurs the usual prescription for which
of the poles are most physically relevant. The parameters of
the pole which has an
energy exactly at the $n\: ^7$Be threshold is displayed in 
Tables~\ref{tab1}-\ref{tab2} because its $E_x$ and $\Gamma$ correspond
most closely to other analyses. There is another nearby pole 
(on the unphysical sheet closest to the physical sheet) with
$E_x=18.73$ MeV, a much larger width 
$\Gamma=640$ keV, $T=1$ and isospin impurity $31$\%. This pole has the opposite
pattern of coupling to the channels: it  couples stronger to $p\: ^7$Li
and weaker to $n\: ^7$Be.

The $1^-1(22)$ resonance has previously only been observed in the
$^7Li(p,\gamma_0)$ reaction~\cite{tunl}. This analysis finds a need to
introduce this resonance with a strong coupling to $p\: ^7$Li and $n\: ^7$Be
in the spin $2$, D-wave. The parameters of $1^-1(22)$ are not strongly 
fixed by this analysis and are hence not displayed.  

\section{Conclusions}

The $^8$Be resonance parameters of most of the resonances up to $26$
MeV are determined. The isospins of the $19$ MeV $J^\pi=1^-$ and the
$4^-$ resonances are determined for the first time to be 0. The $21$
MeV resonance which was previously assigned to possibly have positive
parity is found to be $J^\pi T=3^-0$. The previously known $22$ MeV
$2^+0$ resonance likely splits into two resonances. A new $1^+1$
resonance at $20$ MeV is discovered. The resonance parameters enable
comparison with GFMC and NCSM theory calculations. Two broad
resonances are found, which may not appear in these calculations that
treat the states as bound. These resonances are the extremely broad $2^+0$
resonance at $16$ MeV, whose existence is confirmed, and a very broad
$4^+0$ resonance at $18$ MeV, which is discovered for the first
time. The location of the $T = 1$ resonances is relevant to sorting
out the structure of $^8$Li and $^8$B. Incorporation of the resonance
structure found here in future TUNL evaluations is advocated.

\vspace{0.2cm}

Helpful discussions 
with G.M.~Hale are
gratefully acknowledged. The RESP code for the extraction of S-matrix
poles was written by G.M.~Hale. 
Some of the data analysed was entered by others, including 
G.M.~Hale and A.S.~Johnson~\cite{laur}. 
P. Navr\'{a}til provided detailed calculations
further to Refs.~\cite{navratil04,navratil01}. 
This research is supported by the
Department of Energy under contract W-7405-ENG-36.

\end{document}